\documentclass[11pt,a4paper]{article}

\usepackage{epsfig,amsthm,amscd,amsxtra,amsbsy}
\usepackage{amsmath,amssymb,latexsym, amsfonts}

\begin{document}

\newcommand{\bp}{{\mbox{}\hskip 0.2cm}}
\newcommand{\ba}{\begin{equation}}
\newcommand{\ea}{\end{equation}}

\newtheorem{theorem}{Theorem}[section]
\newtheorem{lemma}{Lemma}[section]
\newtheorem{proposition}{Proposition}[section]
\newtheorem{definition}{Definition}[section]
\newtheorem{assumption}{Assumption}[section]

\textwidth 180mm
\textheight 210mm

\baselineskip 0.6cm
\parskip 0.0cm
\parindent 0.7cm

\pagenumbering{arabic}

\title{\Large\bf Symplectic Schemes for Birkhoffian system \thanks{
\quad Supported by the Special Funds for Major State Basic
Research Projects , G 1999, 032800}}
\author{Hongling Su\quad\quad\quad
 Mengzhao Qin\\
 CAST (World Laboratory), \\
Institute of Computational Mathematics\\
 and Scientific/Engineering Computing, \\
Academy of Mathematics and System Sciences,\\
Chinese Academy of Sciences, Beijing, 100080, China.}
\maketitle
\begin{abstract} A universal symplectic structure for a Newtonian system including nonconservative cases can be constructed in the framework of Birkhoffian generalization of Hamiltonian mechanics. In this paper the symplectic geometry structure of  Birkhoffian system is discussed, then the symplecticity of Birkhoffian phase flow is presented. Based on these properties we give a way to construct symplectic schemes for Birkhoffian systems by using the generating function method.   
\end{abstract}
\vskip 0.3cm
\setcounter{page}{1}
\section{\Large\bf Introduction}
Birkhoffian representation is the generalization of Hamiltonian representation, which can be applied to hadron physics, statistical mechanics, space mechanics, engineering, biophysics and so on\cite{Santilli1,Santilli2}. All conservative or nonconservative, self-adjoint or non-self-adjoint, unconstrained or nonholonomic constrained systems always admit a representation of Birkhoff's equations\cite{Santilli2,Guo}. In last 20 years, many researchers have studied Birkhoffian mechanics and obtained a series of results in integral theory, stability of motion, inverse problem, algebraic and geometric description, and so on.\par
Birkhoff's equations are more complex than Hamilton's equations, so are the studies of their computational method. In the past, there are not any result in the computational methods for Birkhoffian system. The known difference methods are not generally applicable to Birkhoffian system. As the difference scheme to solve Hamiltonian system should be Hamiltonian scheme, the difference scheme to simulate Birkhoffian system should be Birkhoffian scheme, however, the conventional difference schemes such as Euler center scheme, leap-frog scheme and so on are not Birkhoffian scheme, so a method  of how to systematically construct a Birkhoffian scheme is necessary. It is the main context in this paper.\par
The systems both Birkhoffian and Hamiltonian mentioned in this paper are usually finite dimensional situation\cite{Marsden}, in fact, the definition of infinite dimensional Birkhoffian system has not proposed before. Ref. \cite{Santilli2} described the Algebraic and Geometric profiles of the finite dimensional Birkhoffian systems in local coordinates, and general nonautonomous Hamiltonian system  is considered as autonomous Birkhoffian system. Ref. \cite{Feng} developed systematically the symplectic schemes for the standard Hamiltonian system and general Hamiltonian system on the Poisson manifold which belong together the autonomous and semi-autonomous Birkhoffian system. So in this paper, we just discuss the nonautonomous Birkhoffian system in detail.
\par
In section 2, we sketch out Birkhoffian system via variational self-adjoin-\\tness, this sketch shows the relationship between Birkhoffian system and Hamiltonian system more essentially and directly, then we give the basic geometrical properties of Birkhoffian system.
\par
Section 3 extends the definitions of $\widetilde{K}(z)$-Lagrangian submanifold to the one with a parameter $t$ such as $\widetilde{K}(z,t)$-Lagrangian submanifold, then we discuss the relationship between symplectic mappings and gradient mappings.
\par
Section 4 constructs the generating functions for the phase flow of the Birkhoffian system and gives the method to simulate Birkhoffian systems by symplectic schemes of any order. The last section shows an illustrating example. A first-order and a second-order schemes for the system of a linear damped oscillator are given.   
\par
The Einstein's summation convention is used in the next sections. 
\section{\Large\bf Birkhoffian System}
The generalization of Hamilton's equations we shall study is given by
\ba\label{s1.1}
\left(\frac{\partial F_j}{\partial z_i}-\frac{\partial F_i}{\partial z_j}\right)\frac{\text{d}z_i}{\text{d}t}-\left(\frac{\partial B(z,t)}{\partial z_i}+\frac{\partial F(z,t)}{\partial t}\right)=0,
\ea
following the terminology suggested by Santilli(1978)\cite{Santilli2}, we called it as Birkh-\\off's equation or Birkhoffian system. The function $B(z,t)$ is called the Birkhoffian, because of certain physical difference with Hamiltonian which is indicated in Ref. \cite{Santilli2}. $F_i$, $i=1,2,\cdot\cdot\cdot,2n$, are Birkhoffian functions. A representation of Newton's equations via Birkhoff's equation is called a Birkhoffian representation. \par

\begin{definition}\label{def1}
Birkhoff's equations ($\ref{s1.1}$) are called autonomous when the $F_i$ and $B$ functions are free of the time variable, in which case the equations are of the simple form 
\ba\label{def1.1}
K_{ij}(z)\frac{\text{d}z_j}{\text{d}t}-\frac{\partial B(z)}{\partial z_i}=0.
\ea
They are called semi-autonomous when the $F_i$ functions do not depend explicitly on time, in which case we have the more general form
\ba\label{def1.2}
K_{ij}(z)\frac{\text{d}z_j}{\text{d}t}-\frac{\partial B(z,t)}{\partial z_i}=0.
\ea
Birkhoff's equations are called nonautonomous when both the $F_i$ and $B$ functions have  explicit dependence on time, in which case we rewrite
\ba\label{def1.3}
K_{ij}(z,t)\frac{\text{d}z_j}{\text{d}t}-\frac{\partial B(z,t)}{\partial z_i}-\frac{\partial F_i(z,t)}{\partial t}=0.
\ea
Here they all have
\ba\label{def1.4} 
K_{ij}=\frac{\partial F_j}{\partial z_i}-\frac{\partial F_i}{\partial z_j}.
\ea
They are called regular when the functional determinant is not null in the region considered:
\ba\label{def1.5}
\text{det}(K_{ij})(\widetilde\Re)\ne 0,
\ea
otherwise, degenerate.
\end{definition}
Given an arbitrary analytic and regular first-order system
\ba\label{s1.2}
K_{ij}(z,t)\frac{\text{d}z_i}{\text{d}t}+D_i(z,t)=0,\quad\quad\quad i=1,2,...2n,
\ea 
From the point of the inverse variational problem\cite{Atherton}, this system is self-adjoint iff. it satisfies the following condition in $\widetilde\Re^*$, i.e.
\ba\label{s1.3}
\begin{split}
K_{ij}+K_{ji}=0,\\
\frac{\partial K_{ij}}{\partial z_k}+\frac{\partial K_{jk}}{\partial z_i}+\frac{\partial K_{ki}}{\partial z_j}=0,\\
\frac{\partial K_{ij}}{\partial t}=\frac{\partial D_i}{\partial z_j}-\frac{\partial D_j}{\partial z_i},\quad\quad\quad i,j,k=1,2,...2n.
\end{split}
\ea   
We now simply introduce the geometric significance of the condition of variational self-adjointness\cite{Sarlet1,Carinena,Massa,Sarlet2,Morando}. Here the region considered is a star-shaped region $\widetilde\Re^*$ of points of $R\times T^*M$, $T^*M$ the cotangent space of the $M$, $M$ a 2n-dimensional manifold.
\par
Consider first the case for which $K_{ij}=K_{ij}(z)$. Given a symplectic structure written as the 2-form in local coordinates 
\begin{align}\label{s1.4}
\Omega=\sum\limits_{i,j=1}^{2n}K_{ij}(z,t)\text{d}z_i\wedge\text{d}z_j,\quad\quad\quad K_{ij}=-K_{ji},
\end{align} 
one of the fundamental properties of symplectic form ($\ref{s1.4}$) is that $\text{d}\Omega=0$. The geometric significance of the condition of self-adjointness ($\ref{s1.3}$) is then the integrability conditions for 2-form ($\ref{s1.4}$) to be an exact symplectic form coincident with the first two formulas of condition ($\ref{s1.3}$). Because the exact character of two-form ($\ref{s1.4}$) implies following structure
\begin{align}\label{s1.5}
\Omega=\text{d}(F_i\text{d}z_i),
\end{align}
%\ba\label{s1.6}
this geometric property is fully characterized by the first two equations of condition ($\ref{s1.3}$), and we can say that the two-form ($\ref{s1.5}$) describes the geometrical structure of the autonomous case ($\ref{def1.1}$) of the Birkhoff's equations, even it sketches out the geometric structure of the semi-autonomous case. 
\par
For the case of that $K_{ij}=K_{ij}(z,t)$, the full set of condition ($\ref{s1.3}$) must be considered and the corresponding geometric structure can be better expressed by transition of the symplectic geometry on the cotangent bundle $T^*M$ with local coordinates $z_i$ to the contact geometry on the manifold $R\times T^*M$ with local coordinates $\widetilde{z}_i$, $i=0,1,2,\cdot\cdot\cdot,2n$, $\widetilde{z}_0=t^{[1]}$. 
In this case  more general formulations of an exact contact 2-form persist, although it is now referred to as a (2n+1)-dimensional space,
\ba\label{s1.7}
\widehat{\Omega}=\sum\limits_{i,j=0}^{2n}\widehat K_{ij}\text{d}\widetilde{z}_i\wedge\text{d}\widetilde{z}_j=\Omega+2D_i\text{d}z_i\wedge\text{d}t,
\ea  
where
\ba\label{(s1.8)}
\widehat K=\begin{pmatrix}
0 &-D^{\text{T}}\\
D & K
\end{pmatrix},
\ea 
if the contact form is also of the exact type   
\ba\label{s1.9}
\widehat\Omega=\text{d}(\widetilde{F}_i\text{d}\widetilde{z}_i),\quad\quad\widetilde{F}_i=\left\{
\begin{array}{l}
-B\\
F_i
\end{array}
\right.,
\ea
the geometric meaning of the condition of the self-adjointness is then the integrability condition for the exact contact structure ($\ref{s1.9}$).
Here $B$ can be calculated from 
\ba\label{s1.10}
-\frac{\partial B}{\partial z_i}=D_i+\frac{\partial F_i}{\partial t}
\ea
for
\ba\label{s1.11}
\frac{\partial}{\partial z_j}\left(D_i+\frac{\partial F_i}{\partial t}\right)=\frac{\partial}{\partial z_i}\left(D_j+\frac{\partial F_j}{\partial t}\right).
\ea
\par
All the above discussion can be expressed via the following property.
\begin{proposition}\label{prop1}
($\text{Self$-$Adjointness of Birkhoffian System}$). Necessary a-\\nd sufficient condition for a general nonautonomous first-order system given \\as above to be self-adjoint in $\widetilde\Re^{*}$ of points of  $R\times T_*R^{2n}$ is that it is of the Birkhoffian type, i.e.,
\ba\label{prop1.1}
K_{ij}(z,t)\frac{\text{d}z_i}{\text{d}t}+D_i(z,t)=(\frac{\partial F_j}{\partial z_i}-\frac{\partial F_i}{\partial z_j})\frac{\text{d}z_i}{\text{d}t}-(\bigtriangledown B(z,t)+\frac{\partial F(z,t)}{\partial t}).
\ea  
\end{proposition}
The functions $F_i$ and $B$ can be calculated according to the rules\cite{Atherton}
\ba\label{s1.12}
F_i=\frac{1}{2}\int\limits_0^1z_j\cdot K_{ji}(\lambda z,t)\text{d}\lambda,
\ea
\ba\label{s1.13}
B=\int\limits_0^1 z_i\cdot(D_i+\frac{\partial F_i}{\partial t})(\lambda z,t)\text{d}\lambda.
\ea
\par
As well as Hamiltonian system, Birkhoffian system provides a symbiosis among variational principle, Lie's algebra and symplectic geometry.\cite{Feynman,Santilli1,Santilli2,Arnold}
Obviously both standard Hamilton's equations and general Hamilton's equations on Poisson manifold  are recovered from Birkhoff's equations as in the particular cases of autonomous and semi-autonomous Birkhoffian representations. 

\par 
Due to the self-adjointness of Birkhoff's equations, the phase flow of the system ($\ref{prop1.1}$) conserves the symplecticity, then we get
\begin{align}\label{s1.14}
\frac{\text{d}}{\text{d}t}\Omega =\frac{\text{d}}{\text{d}t}(K_{ij}\text{d}z_i\wedge\text{d}z_j)=0.
\end{align} 
It means that if we denote the phase flow of the equations ($\ref{prop1.1}$) with $(\widehat{z},\widehat{t})$, then
\ba\label{s1.15}
K_{ij}(\widehat{z},\widehat{t})\text{d}\widehat{z}_i\wedge\text{d}\widehat{z}_j=K_{ij}(z,t)\text{d}z_i\wedge\text{d}z_j,
\ea
or
\ba\label{s1.16}
{\frac{\partial\widehat{z}}{\partial z}}^{\text{T}}K(\widehat{z},\widehat{t})\frac{\partial\widehat{z}}{\partial z}=K(z,t).
\ea
In the next sections we will  construct the algorithm preserving this geometric property of the phase flow in discrete space.  

\section{\Large\bf Generating Functions for $K(z,t)$-Symplectic Ma\\ppings}
In this section we consider general $k(z,t)$-symplectic mappings and their relationship with the gradient mappings and their generating functions.  
\begin{definition}\label{def2}
Denote
\ba\label{def2.1}
\begin{split}
J_{2n}=
\begin{pmatrix}
0
& I_n\\
-I_n
& 0
\end{pmatrix},
\quad\quad\quad
J_{4n}=
\begin{pmatrix}
0
& I_{2n}\\
-I_{2n}
& 0
\end{pmatrix},\\
\widetilde{J}_{4n}=
\begin{pmatrix}
J_{2n}
& 0\\
0
& -J_{2n}
\end{pmatrix},\quad\quad\quad
\widetilde{K}(\widehat{z},z,t,t_0)=
\begin{pmatrix}
K(\widehat{z},t)
& 0\\
0
& -K(z,t_0)
\end{pmatrix}.
\end{split}
\ea
\par
A $2n$-dimensional submanifold $L\subset R^{4n}$
\ba\label{def2.2}
L=\left\{
\begin{pmatrix}
&\widehat{z}&\\
&z&
\end{pmatrix}
\in R^{4n}|z=z(x,t_0),\widehat{z}=\widehat{z}(x,t),x\in U\subset R^{2n},\text{open set}\right\}
\ea
is a ${J}_{4n}$- or $\widetilde{J}_{4n}$- or $\widetilde{K}(\widehat{z},z,t,t_0)$-Lagrangian submanifold if
\ba\label{def2.3}
(T_xL)^{\text{T}}{J}_{4n}(T_xL)=0
\ea
or
\ba\label{def2.4}
(T_xL)^{\text{T}}\widetilde{J}_{4n}(T_xL)=0
\ea
or
\ba\label{def2.5}
(T_xL)^{\text{T}}\widetilde{K}(\widehat{z},z,t,t_0)(T_xL)=0,
\ea
where $T_xL$ is the tangent space to $L$ at $x$.
\par
A mapping with parameters $t$ and $t_0$ is $z\longrightarrow\widehat{z}=g(z,t,t_0): R^{2n}\longrightarrow R^{2n}$ which is called a canonical map or a gradient map or a K(z,t)-symplectic map if its graph 
\ba\label{s2.1}
\Gamma_g=\left\{
\begin{pmatrix}
&\widehat{z}&\\
&z&
\end{pmatrix}
\in R^{4n}|\widehat{z}=g(z,t,t_0),z=z\in R^{2n}\right\}
\ea
is a ${J}_{4n}$- or $\widetilde{J}_{4n}$- or $\widetilde{K}(\widehat{z},z,t,t_0)$-Lagrangian submanifold.
\end{definition}

\begin{definition}\label{def3}
A differentiable mapping $g: M\rightarrow M$ is $K(z,t)$-symplectic,\\if 
\ba\label{def3.1}
{\frac{\partial g}{\partial z}}^{\text{T}}K(g(z,t,t_0),t)\frac{\partial g}{\partial z}=K(z,t_0).
\ea
A difference scheme approximating the Birkhoff's system ($\ref{prop1.1}$)
\ba\label{def3.2}
z^{k+1}=g^k(z^k,t_k+\tau,t_k),\quad\quad\quad k\geqslant 0,
\ea
either explicit scheme or worked out from a inexplicit scheme, is called a K-symplectic scheme, when $g^k$ is K-symplectic for every $k\geqslant 0$, i.e.
\ba\label{def3.3}
{\frac{\partial g^k}{\partial z^k}}^{\text{T}}K(z^{k+1},t^{k+1})\frac{\partial g
^k}{\partial z^k}=K(z^k,t^k).
\ea
\end{definition}
The graph of the phase flow of the Birkhoffian system ($\ref{s1.1}$) is $g^t(z,t_0)=g(z,t,t_0)$ which is a $\widetilde{K}(\widehat{z},z,t,t_0)$-Lagrangian submanifold for
\ba\label{s2.2}
g^{t}_z(z,t_0)^{\text{T}}K(g^{t}(z,t_0),t)g^{t}_z(z,t_0)=K(z,t_0).
\ea
Similarly the graph of the phase flow of standard Hamiltonian system is a $\widetilde{J}_{4n}$-Lagrangian submanifold.
\par
Define nonlinear transformation with two parameters $t$ and $t_0$ from $R^{4n}$ to itself,
$$
\alpha(t,t_0):
\begin{pmatrix}
&\widehat{z}&\\
& z &
\end{pmatrix}
\longrightarrow
\begin{pmatrix}
&\widehat{w}&\\
& w &
\end{pmatrix}
=
\begin{pmatrix}
&\alpha_1(\widehat{z},z,t,t_0)&\\
& \alpha_2(\widehat{z},z,t,t_0)&
\end{pmatrix},
$$
\ba\label{s2.3}
\alpha^{-1}(t,t_0):
\begin{pmatrix}
&\widehat{w}&\\
&w&
\end{pmatrix}
\longrightarrow
\begin{pmatrix}
&\widehat{z}&\\
&z&
\end{pmatrix}
=\begin{pmatrix}
&\alpha^1(\widehat{w},w,t,t_0)&\\
&\alpha^2(\widehat{w},w,t,t_0)&
\end{pmatrix}.
\ea
Denote
$$
\alpha_*(\widehat{z},z,t,t_0)=
\begin{pmatrix}
A_\alpha
&B_\alpha\\
C_\alpha
&D_\alpha
\end{pmatrix},
$$
\ba\label{s2.4}
\alpha_*^{-1}(\widehat{w},w,t,t_0)=
\begin{pmatrix}
A^\alpha
&B^\alpha\\
C^\alpha
&D^\alpha
\end{pmatrix}.
\ea
$ \alpha_* $ 
is the Jacobian of $\alpha$. Let $\alpha$ be a diffeomorphism from $R^{4n}$ to itself, then it follows that $\alpha$ carries every $\widetilde{K}$-Lagrangian submanifold into a $J_{4n}$-Lagrangian submanifold, iff.
\ba\label{s2.5}
\alpha^{\text{T}}_*J_{4n}\alpha_*=\widetilde{K},
\ea
i.e.,
\ba\label{s2.6}
\begin{pmatrix}
A_\alpha
&B_\alpha\\
C_\alpha
&D_\alpha
\end{pmatrix}^{\text{T}}
\begin{pmatrix}
J_{2n} &0\\
0 &-J_{2n}
\end{pmatrix}
\begin{pmatrix}
A_\alpha
&B_\alpha\\
C_\alpha&D_\alpha
\end{pmatrix}=
\begin{pmatrix}
K(\widehat{z},t)
& 0\\
0
& -K(z,t_0)
\end{pmatrix}.
\ea
Conversely
$\alpha^{-1}$ carries every $J_{4n}$-Lagrangian submanifold into a $\widetilde{K}$-Lagrang-\newline ian submanifold.

\begin{proposition}\label{prop2}
Let $M\in R^{2n\times 2n}$, $\alpha$ be defined as above. Define a fractional transformation
\ba\label{prop2.1}
\begin{split}
\sigma_{\alpha}:&\mathcal{M}\longrightarrow\mathcal{M}\\
&M\longrightarrow N=\sigma_\alpha(M)=(A_\alpha M+B_\alpha)(C_\alpha M+D_\alpha)^{-1}
\end{split}
\ea
under the transversality condition 
\ba\label{prop2.2}
|C_\alpha M+D_\alpha|\ne 0.
\ea
Then the following four conditions are equivalent mutually:
\ba
\begin{split}
&|C_\alpha M+D_\alpha|\ne 0,\\
&|MC^\alpha-A^\alpha|\ne 0,\\
&|C^\alpha N+D^\alpha|\ne 0,\\
&|NC_\alpha-A_\alpha|\ne 0.
\end{split}
\ea 
\end{proposition}
The proof is direct and simple, so it is omitted here.
\begin{theorem}\label{th0}
Let $\alpha$ be defined as above. Let $z\longrightarrow\widehat{z}=g(z,t,t_0)$ be a K(z,t)-symplectic mapping in some neighborhood $\widetilde{\mathcal{R}}$ of $R^{2n}$ with Jacobian $g_z(z,t,t_0)=M(z,t,t_0)$. If M satisfies the transversality condition in $\widetilde{\mathcal{R}}$ 
\ba\label{th0.1}
|C_\alpha(g(z,t,t_0),z,t,t_0)M(z,t,t_0)+D_\alpha(g(z,t,t_0),z,t,t_0)|\ne 0,
\ea 
then there exists uniquely in $\widetilde{\mathcal{R}}$ a gradient mapping $w\longrightarrow\widehat{w}=f(w,t,t_0)$ with Jacobian $f_w(w,t,t_0)=N(w,t,t_0)$ and a scalar function-generating function-$\phi(w,t,t_0)$ such that
\ba\label{th0.2}
f(w,t,t_0)=\phi_w(w,t,t_0),
\ea
\ba\label{th0.3}
\begin{split}
\alpha_1(g(z,t,t_0),z,t,t_0)&=f(\alpha_2(g(z,t,t_0),z,t,t_0),t,t_0)\\
&=\phi_w(\alpha_2(g(z,t,t_0),z,t,t_0),t,t_0), \\
\end{split}
\ea
 identically in z and t,

\ba\label{th0.4}
N=(A_\alpha M+B_\alpha)(C_\alpha M+D_\alpha)^{-1},
\ea
\ba\label{th0.5}
M=(A^\alpha N+B^\alpha)(C^\alpha N+D^\alpha)^{-1}.
\ea
\end{theorem}
{\bf Proof.} Under the transformation $\alpha$, the image of the graph $\Gamma_g$ is
\ba\label{th0.6}
\begin{split}
\alpha(\Gamma_g)=\left\{
\begin{pmatrix}
&\widehat{w}&\\
&w& 
\end{pmatrix}\right.
\in R^{4n}|\widehat{w}&=\alpha_1(g(z,t,t_0),z,t,t_0),\\
w&=\alpha_2(g(z,t,t_0),z,t,t_0)\left.\right\}.
\end{split}
\ea
By the inequality ($\ref{th0.1}$),
\ba\label{th0.7}
\frac{\partial w}{\partial z}=\frac{\partial \alpha_2}{\partial \widehat{z}}\cdot\frac{\partial\widehat{z}}{\partial z}+\frac{\partial \alpha_2}{\partial z}=C_\alpha M+D_\alpha\not=0,
\ea
so $w=\alpha_2(g(z,t,t_0),z,t,t_0)$ is invertible, the inverse function is denoted by
$z=z(w,t,t_0)$.
Set
\begin{align}\label{th0.8}
\widehat{w}=f(w,t,t_0)=\alpha_1(g(z,t,t_0),z,t,t_0)|_{z=z(w,t,t_0)},
\end{align}
then
\ba\label{th0.9}
N=\frac{\partial f}{\partial w}=(\frac{\partial\alpha_1}{\partial\widehat{z}}\frac{\partial g}{\partial z}+\frac{\partial\alpha_1}{\partial z})(\frac{\partial z}{\partial w})=(A_\alpha M+B_\alpha)(C_\alpha M+D_\alpha)^{-1}.
\ea
Notice that the tangent space to $\alpha(\Gamma_g)$ at $z$ is
\ba\label{th0.10}
T_z(\alpha(\Gamma_g))=
\begin{pmatrix}
&\frac{\partial\widehat{w}}{\partial z}&\\
&\frac{\partial w}{\partial z}&
\end{pmatrix}=
\begin{pmatrix}
&A_\alpha M+B_\alpha&\\
&C_\alpha M+D_\alpha&
\end{pmatrix}.
\ea
we have a conclusion that $\alpha\left(\Gamma_g\right)$ is a $J_{4n}$-Lagrangian submanifold for
\begin{align}\label{th0.11}
\begin{split}
& T_z(\alpha(\Gamma_g))^{\text{T}}J_{4n}T_z(\alpha(\Gamma_g))\\
&=\left((A_\alpha M+B_\alpha)^{\text{T}},(C_\alpha M+D_\alpha)^{\text{T}}\right)
J_{4n}
\begin{pmatrix}
&A_\alpha M+B_\alpha&\\
&C_\alpha M+D_\alpha&
\end{pmatrix}\\
&=\left(M^{\text{T}},I\right)\alpha^{\text{T}}_*J_{4n}\alpha_*
\begin{pmatrix}
& M &\\
& I &
\end{pmatrix}\\
&=\left(M^{\text{T}},I\right)\widetilde{K}
\begin{pmatrix}
& M &\\
& I &
\end{pmatrix}
=0.
\end{split}
\end{align}
So
\ba\label{th0.12}
(A_\alpha M+B_\alpha)^{\text{T}}(C_\alpha M+D_\alpha)-(C_\alpha M+D_\alpha)^{\text{T}}(A_\alpha M+B_\alpha)=0,
\ea 
i.e., $N=(A_\alpha M+B_\alpha)(C_\alpha M+D_\alpha)^{-1}$ is symmetric. It implies that $\widehat{w}=f(w,t,t_0)$ is a gradient mapping. By the Poincar$\acute{e}$ lemma, there is a scalar function $\phi(w,t,t_0)$ such that
\ba\label{th0.13}
f(w,t,t_0)=\phi_w(w,t,t_0).
\ea
The equation ($\ref{th0.3}$) follows from the construction of $f(w,t,t_0)$ and $z(w,t,t_0)$.\\Since $z(w,t,t_0)\circ\alpha_2(g(z,t,t_0),z,t,t_0)\equiv z$, so substituting $w=\alpha_2(g(z,t,t_0),\\z,t,t_0)$ in the equations ($\ref{th0.8}$) and ($\ref{th0.13}$), we get the equation ($\ref{th0.3}$).
\par
\begin{proposition}\label{prop3}
$f(w,t,t_0)$ obtained in Theorem $\ref{th0}$ is also the solution of the following implicit equation
\ba\label{prop3.1}
\alpha^1(f(w,t,t_0),w,t,t_0)=g(\alpha^2(f(w,t,t_0),w,t,t_0),t,t_0).
\ea
\end{proposition}
The proof is similar to the above proof.
\begin{theorem}\label{th1}
Let $\alpha$ be  defined as in Theorem $\ref{th0}$. Let $w\longrightarrow\widehat{w}=f(w,t,t_0)$ be a gradient mapping in some neighborhood $\widetilde{\mathcal{R}}$ of $R^{2n}$ with Jacobian $f_w(w,t,t_0)=N(w,t,t_0)$. If in $\widetilde{\mathcal{R}}$, N satisfies the condition
\ba\label{th1.1}
|C^\alpha(f(w,t,t_0),w,t,t_0)N(w,t,t_0)+D^\alpha(f(w,t,t_0),w,t,t_0)|\ne0,
\ea 
then there exits uniquely in $\widetilde{\mathcal{R}}$, a K(z,t)-symplectic mapping $z\longrightarrow\widehat{z}=g(z,t,t_0)$ with Jacobian $g(z,t,t_0)=M(z,t,t_0)$ such that
\ba\label{th1.2}
\begin{split}
\alpha^1(f(w,t,t_0),w,t,t_0)=g(\alpha^2(f(w,t,t_0),w,t,t_0),t,t_0),\\
M=(A^\alpha N+B^\alpha)(C^\alpha N+D^\alpha)^{-1},\\
N=(A_\alpha M+B_\alpha)(C_\alpha M+D_\alpha)^{-1}.
\end{split}
\ea\\
Similarly to Proposition $\ref{prop3}$, $g(z,t,t_0)$ is the solution of the implicit equation
\ba\label{th1.3}
\alpha_1(g(z,t,t_0),z,t,t_0)=f(\alpha_2(g(z,t,t_0),z,t,t_0),z,t,t_0).
\ea
\end{theorem}
The proof is similar to that of Theorem $\ref{th0}$ and is omitted here.
\section{\Large\bf Symplectic Difference Schemes for Birkhoff's Eq\\uations}
In  Section 2  it is indicated that for a general Birkhoff's system, there is a common property that its phase flow is symplectic. Through the result in the proceeded section, we construct symplectic schemes for  Birkhoff's system by approximating the generating functions.
\par
The Birkhoff's phase flow is denoted by $g^{t}(z,t_0)$, it is a one-parameter group of $K(z,t)$-symplectic mappings at least local in $z$ and $t$, i.e.,
\ba
g^{t_0}=\text{identity},\quad\quad\quad g^{t_1+t_2}=g^{t_1}\circ g^{t_2},
\ea
here $z$ is taken as an initial value when $t=t_0$, and $\widehat{z}(z,t,t_0)=g^{t}(z,t_0)=g(t;z,t_0)$ is the solution of the Birkhoffian system ($\ref{prop1.1}$).
\begin{theorem}\label{th2}
Let $\alpha$  be defined as above. Let $z\rightarrow \widehat{z}=g^{t}(z,t_0)$ be the phase flow of the Birkhoff's system ($\ref{prop1.1}$), $M(t;z,t_0)=g_z(t;z,t_0)$ is its Jacobian. At some initial point $z$, i.e., $t=t_0$, $\widehat{z}=z$, if 
\ba\label{th2.1}
|C_\alpha(z,z,t_0,t_0)+D_\alpha(z,z,t_0,t_0)|\ne0,
\ea
then there exists, for sufficiently small $|t-t_0|$ and in some neighborhood o-\\f $z\in R^{2n}$, a gradient mapping $w\rightarrow\widehat{w}=f(w,t,t_0)$ with Jacobian $f_w(w,t,t_0)\\=N(w,t,t_0)$ symmetric and a scalar function-generating function-$\phi(w,t,t_0)$ such that
\ba\label{th2.2}
 f(w,t,t_0)=\phi_w(w,t,t_0),
\ea
\ba\label{th2.3}
\frac{\partial}{\partial t}\phi_w(w,t,t_0)=\mathcal{A}(\phi_w(w,t,t_0),w,\phi_{ww}(w,t,t_0),t,t_0),
\ea
\ba\label{th2.4}
\mathcal{A}(\widehat{w},w,\frac{\partial\widehat{w}}{\partial w},t,t_0)=\Bar{\mathcal{A}}(\widehat{z}(\widehat{w},w,t,t_0),z(\widehat{w},w,t,t_0),\frac{\partial\widehat{w}}{\partial w},t,t_0),
\ea
\ba\label{th2.5}
\begin{split}
\Bar{\mathcal{A}}(\widehat{z},z,\frac{\partial\widehat{w}}{\partial w},t,t_0) &=\frac{\text{d}}{\text{d}t}\widehat{w}(\widehat{z},z,t,t_0)-\frac{\partial\widehat{w}}{\partial w}\frac{\text{d}}{\text{d}t}w(\widehat{z},z,t,t_0)\\
&=(A_\alpha-\frac{\partial\widehat{w}}{\partial w}C_\alpha)K^{-1}D(\widehat{z},t)+\frac{\partial\alpha_1}{\partial t}-\frac{\partial\widehat{w}}{\partial w}\frac{\partial\alpha_2}{\partial t},
\end{split}
\ea
\ba\label{th2.6}
\begin{split}
\alpha_1(g(t;z,t_0),z,t,t_0)&=f(\alpha_2(g(t;z,t_0),z,t,t_0),t,t_0)\\
&=\phi_w(\alpha_2(g(t;z,t_0),z,t,t_0),t,t_0), \\
\end{split}
\ea
 identically in z and t,
\ba\label{th2.7}
\begin{split}
&N=\sigma_\alpha(M)=(A_\alpha M+B_\alpha)(C_\alpha M+D_\alpha)^{-1},\\
& M=\sigma_{\alpha^{-1}}=(A^\alpha N+B^\alpha)(C^\alpha N+D^\alpha)^{-1}.
\end{split}
\ea
\end{theorem}
{\bf Proof.} $M(t;z,t_0)$ is differentiable with respect to $z$ and $t$. Condition ($\ref{th2.1}$) guarantees that for sufficiently small $|t-t_0|$ and in some neighborhood $\widehat{z}$ of $z\in R^{2n}$, there is
\ba\label{th2.8}
|C_{\alpha}(\widehat{z},z,t,t_0)M(t;z,t_0)+D_{\alpha}(\widehat{z},z,t,t_0)|\not=0.
\ea
Additionally the Birkhoffian phase flow is a symplectic mapping, therefore by Theorem $\ref{th0}$ there exists a time-dependent gradient map $\widehat{w}=f(w,t,t_0)$ and there is  a scalar function $\phi(w,t,t_0)$, such that
\ba\label{th2.9}
f(w,t,t_0)=\phi_w(w,t,t_0).
\ea
\ba\label{th2.10}
\frac{\partial f(w,t,t_0)}{\partial t}=\frac{\partial\phi_w(w,t,t_0)}{\partial t}
\ea
\par
Notice that $\widehat{z}=g(t;z,t_0)$ is the solution of the following initial-value problem
\ba\label{th2.11}
\left\{ 
\begin{array}{l}
\frac{\text{d}\widehat{z}}{\text{d}t}=K^{-1}(\widehat{z},t)(\bigtriangledown B+\frac{\partial F}{\partial t})(\widehat{z},t)\\
\widehat{z}|_{t=t_0}=z
\end{array},\right.
\ea
therefore from the equations
\ba\label{th2.12}
\widehat{w}=\alpha_1(\widehat{z},z,t,t_0),\quad\quad\quad w=\alpha_2({\widehat{z},z,t,t_0}),
\ea
it follows that
\ba\label{th2.13}
\begin{split}
\frac{\text{d}\widehat{w}}{\text{d}t}&=\frac{\partial\widehat{w}}{\partial\widehat{z}}\cdot\frac{\text{d}\widehat{z}}{\text{d}t}+\frac{\partial}{\partial t}\alpha_1(\widehat{z},z,t,t_0)\\
&=A_\alpha K^{-1}(\bigtriangledown B+\frac{\partial F}{\partial t})+\frac{\partial \alpha_1}{\partial t},\\
\frac{\text{d}w}{\text{d}t}&=C_\alpha K^{-1}(\bigtriangledown B+\frac{\partial F}{\partial t})+\frac{\partial \alpha_2}{\partial t},
\end{split}
\ea
so
\ba\label{th2.14}
\begin{split}
\frac{\partial\widehat{w}}{\partial t}&=\frac{\text{d}\widehat{w}}{\text{d}t}-\frac{\partial\widehat{w}}{\partial t}\frac{\text{d}w}{\text{d}t}\\
&=(A_\alpha-\frac{\partial\widehat{w}}{\partial t}C_\alpha)K^{-1}(\bigtriangledown B+\frac{\partial F}{\partial t})+\frac{\partial \alpha_1}{\partial t}-\frac{\partial\widehat{w}}{\partial w}\frac{\partial \alpha_2}{\partial t}.
\end{split}
\ea
Since $\frac{\partial\widehat{w}}{\partial w}\ne 0$, so $w=w(\widehat{w},t)$ exists and it is solvable, But it can not be solved directly from the transformation $\alpha$ and $\alpha^{-1}$.
Set
\ba\label{th2.15}
\Bar{\mathcal{A}}(\widehat{z},z,\frac{\partial\widehat{w}}{\partial w},t,t_0)=\frac{\partial\widehat{w}}{\partial t},
\ea 
then we denote
\ba\label{th2.16}
\mathcal{A}(\widehat{w},w,\frac{\partial\widehat{w}}{\partial w},t,t_0)=\Bar{\mathcal{A}}(\widehat{z}(\widehat{w},w,t,t_0),z(\widehat{w},w,t,t_0),\frac{\partial\widehat{w}}{\partial w},t,t_0),
\ea
so from the equation ($\ref{th2.10}$), we get 
\ba\label{th2.17}
\frac{\partial}{\partial t}\phi_w(w,t,t_0)=\mathcal{A}(\phi_w,w,\phi_{ww},t,t_0).
\ea
Q.E.D..\par
The above theorem is given for the nonautonomous Birkhoffian system, which is geometrically different from the autonomous or semi-autonomous Birkhoffian system and we explained this point in Section 2. According to Ref. \cite{Feng}, we can easily construct any order symplectic difference schemes for the later two. Because of the simplicity of the ordinary geometry structure in autonomous or semi-autonomous case, the transformation $\alpha$ we need in these cases is free of parameter $t$, and the corresponding Birkhoffian is completely Hamiltonian, so 

\ba
\begin{split}
\frac{\partial\widehat{w}}{\partial t}&=\frac{\text{d}\widehat{w}}{\text{d}t}-\frac{\partial\widehat{w}}{\partial t}\frac{\text{d}w}{\text{d}t}\\
&=(A_\alpha-\frac{\partial\widehat{w}}{\partial t}C_\alpha)K^{-1}\bigtriangledown B\\
&=-(B^{\alpha\text{T}}+(\frac{\partial\widehat{w}}{\partial w}^{\text{T}})A^{\alpha\text{T}})\bigtriangledown_zB\\
&=-B_w(\widehat{z}(\widehat{w},w))\quad(\text{or} =-B_w(\widehat{z}(\widehat{w},w),t)),
\end{split}
\ea     
resonantly we have the Hamilton-Jacobi equation
\ba
\frac{\partial\phi(w,t)}{\partial t}=-B(\widehat{z}(\phi_w,w)),
\ea
or for semi-autonomous cases there is
\ba
\frac{\partial\phi(w,t,t_0)}{\partial t}=-B(\widehat{z}(\phi_w,w),t). 
\ea 
\par
We describe the $k$-th order total derivative of $\mathcal{A}$ with respect to $t$ as
\ba
\begin{split}
D_t^k\mathcal{A}&=\partial_{\phi_w}\mathcal{A}(\sum\limits_{i=0}^{\infty}(t-t_0)^i\phi_w^{(k+i)})+\partial_{\phi_{ww}}\mathcal{A}(\sum\limits_{i=0}^{\infty}(t-t_0)^i\phi_{ww}^{(k+i)})+\\
&\partial_t\partial_{\phi_w}\mathcal{A}(\sum\limits_{i=0}^{\infty}(t-t_0)^i\phi_w^{(k-1+i)}+\partial_t\partial_{\phi_{ww}}\mathcal{A}(\sum\limits_{i=0}^{\infty}(t-t_0)^i\phi_{ww}^{(k-1+i)})+\\
&\sum\limits_{m=0}^{k}C_k^m\sum\limits_{n=1}^{k-m}C_{k-m}^n\sum\limits_{l=1}^{k-m-n}\sum\limits_{\stackrel{h_1+\cdot\cdot\cdot+h_n+}{j_1+\cdot\cdot\cdot+j_l=k-m}}\partial_{\phi_w}^n\partial_{\phi_{ww}}^l\partial_t^m\mathcal{A}\\
&(\sum\limits_{i=0}^{\infty}(t-t_0)^i\phi_w^{(h_1+i)},\cdot\cdot\cdot,\sum\limits_{i=0}^{\infty}(t-t_0)^i\phi_w^{(h_n+i)},\\
&\sum\limits_{i=0}^{\infty}(t-t_0)^i\phi_{ww}^{(j_1+i)},\cdot\cdot\cdot,\sum\limits_{i=0}^{\infty}(t-t_0)^i\phi_w^{(j_l+i)}),
\end{split}
\ea
then at the point of $t=t_0$, the total derivative of $\mathcal{A}$ is as
\ba
\begin{split}
D_t^k\mathcal{A}(&\phi_w^{(0)},w,\phi_{ww}^{(0)},t_0,t_0)=\partial_{\phi_w}\mathcal{A}\phi_w^{(k)}+\partial_{\phi_{ww}}\mathcal{A}\phi_{ww}^{(k)}+\\
&\partial_t\partial_{\phi_w}\mathcal{A}\phi_w^{(k-1)}+\partial_t\partial_{\phi_ww}\mathcal{A}\phi_{ww}^{(k-1)}+\\
&\sum\limits_{m=0}^{k}C_k^m\sum\limits_{n=1}^{k-m}C_{k-m}^n\sum\limits_{l=1}^{k-m-n}\sum\limits_{\stackrel{h_1+\cdot\cdot\cdot+h_n+}{j_1+\cdot\cdot\cdot+j_l=k-m}}\partial_{\phi_w}^n\partial_{\phi_{ww}}^l\partial_t^m\\
&\mathcal{A}(\phi_w^{(0)},w,\phi_{ww}^{(0)},t_0,t_0)(\phi_w^{(h_1)},\cdot\cdot\cdot,\phi_w^{(h_l)},\phi_{ww}^{(j_1)},\cdot\cdot\cdot,\phi_{ww}^{(j_n)}).
\end{split}
\ea
\begin{theorem}\label{th3}
Let $\mathcal{A}$ and $\alpha$ be analytic. Then the generating function $\phi_{w\alpha,\mathcal{A}}(w,t,t_0)=\phi_w(w,t,t_0)$ can be expanded as a convergent power series in t for sufficiently small $|t-t_0|$
\ba\label{th3.1}
\phi_w(w,t,t_0)=\sum\limits_{k=0}^\infty(t-t_0)^k\phi_w^{(k)}(w,t_0),
\ea
and $\phi_w^{(k)},\;k\geqslant 0$, can be recursively determined by the following equations
\ba\label{th3.2}
\phi_w^{(0)}(w,t_0)=f(w,t_0,t_0),
\ea
\ba\label{th3.3}
\phi_w^{(1)}(w,t_0)=\mathcal{A}(\phi_w^{(0)},w,\phi_{ww}^{(0)},t_0,t_0),
\ea
\ba\label{th3.4}
\phi_w^{k+1}(w,t_0)=\frac{1}{(k+1)!}D_t^k\mathcal{A}(\phi_w^{(0)},w,\phi_{ww}^{(0)},t_0,t_0).
\ea
\end{theorem}
{\bf Proof.} Differentiating the equation ($\ref{th3.1}$) with respect to
$w$ and $t$, we get   
\ba\label{th3.6}
\phi_{ww}(w,t,t_0)=\sum\limits_{k=0}^{\infty}(t-t_0)^k\phi_{ww}^{(k)}(w,t_0),
\ea 
and
\ba\label{th3.7}
\frac{\partial}{\partial t}\phi_w(w,t,t_0)=\sum\limits_{k=0}^{\infty}(k+1)(t-t_0)^k\phi_w^{(k+1)}(w,t_0).
\ea
By the equation ($\ref{th2.2}$),
\ba\label{th3.8}
\phi_w^0(w,t_0)=\phi_w(w,t_0,t_0)=f(w,t_0,t_0).
\ea 
Substituting the equation ($\ref{th3.6}$) in $\mathcal{A}(\widehat{w},w,\frac{\partial\widehat{w}}{\partial w},t,t_0)$, and expanding $\mathcal{A}$ in $t=t_0$, we get
\ba\label{th3.9}
\begin{split}
\mathcal{A}(\phi_w,w,\phi_{ww},t,t_0)&=\mathcal{A}(f(w,t_0,t_0),w,f_w(w,t_0,t_0),t_0,t_0)+\\
&\sum\limits_{k=1}^{\infty}\frac{1}{k!}(t-t_0)^kD_t^k\mathcal{A}(\phi_w^{(0)},w,\phi_{ww}^{(0)},t_0,t_0).
\end{split}
\ea
Using the equation ($\ref{th2.3}$) and comparing the equation ($\ref{th3.9}$) with the equation ($\ref{th3.7}$), we get the equations ($\ref{th3.3}$) and ($\ref{th3.4}$).
\par
For the autonomous and semi-autonomous the difference is that $\mathcal{A}$ is replaced with Bikhoffian $B$, which makes the expanding of generating function $\phi$ much more easy. 
\par
Through Theorem ($\ref{th2}$) and ($\ref{th3}$), the relationship between the Birkhoffian phase flow and the generating function $\phi(w,t,t_0)$ is established. With the aid of this result, we can directly construct $k(z,t)$-symplectic difference schemes.

\begin{theorem}\label{th4}
Let $\alpha$ be given as Theorem ($\ref{th2}$), and $\mathcal{A}$ analytic. For sufficiently small $\tau>0$ as the time-step. Take
\ba\label{th4.1}
\psi_w^{(m)}(w,t_0+\tau,t_0)=\sum\limits_{i=0}^m\tau^i\phi_w^{(i)}(w,t_0),\quad\quad\quad m=1,2,\cdot\cdot\cdot,
\ea
where $\phi_w^{(i)}$ are determined by equations ($\ref{th3.2}$)-($\ref{th3.4}$). Then $\psi_w^{(m)}(w,t_0+\tau,t_0)$ defines a K(z,t)-symplectic difference scheme $z=z^k\longrightarrow z^{k+1}=\widehat{z}$,
\ba\label{th4.2}
\alpha_1(z^{k+1},z^k,t_{k+1},t_k)=\psi_w^{(m)}(\alpha_2(z^{k+1},z^k,t_{k+1},t_k),t_{k+1},t_k)
\ea
of m-th order of accuracy.
\end{theorem}
{\bf Proof.} Because of that
\ba\label{th4.3}
|C_\alpha(z,z,t_0,t_0)+D_\alpha(z,z,t_0,t_0)|\ne0,
\ea
so by Proposition $\ref{prop2}$, $|C^\alpha N+D^\alpha|\ne 0$, where $N=\phi_{ww}(w_0,t_0,t_0)=\psi_{ww}^{(m)}(w_0,t_0,t_0)$, and $w_0=\alpha(z,z,t_0,t_0)$. Thus for sufficiently small $\tau$ and in some neighborhood of $w_0$, there exists
\ba\label{th4.4}
|C^\alpha N^{(m)}(w,t_0+\tau,t_0)+D^\alpha|\ne 0,
\ea
where 
\ba\label{th4.5}
 N^{(m)}(w,t_0+\tau,t_0)=\psi_{ww}^{(m)}(w,t_0+\tau,t_0).
\ea
\par
By Theorem $\ref{th1}$, $\psi_w^{(m)}(w,t_0+\tau,t_0)$ defines a $K(z,t)$-symplectic mapping which is expressed in the equation ($\ref{th1.3}$). Therefore the equation ($\ref{th4.2}$) determines a $m$-th order $K(z,t)$-symplectic difference scheme for the Birkhoffian system ($\ref{prop1.1}$).
\section{Example}
In this section we give an example to illustrate how to get a symplectic-structure-preserving scheme for a non-conservative system which can be expressed in Birkhoffian representation.
\par
Consider the system of a linear damped oscillator
\begin{align}\label{s5.1}
\ddot{r}+\nu\dot{r}+r=0. 
\end{align}
We introduce a gradient function $p$ satisfying $p=\dot r$, then a Birkhoffian representation of this system is given by 
\begin{align}\label{s5.2}
\begin{pmatrix}
0&-\text{e}^{\nu t}&\\
\text{e}^{\nu t}&0&
\end{pmatrix}\begin{pmatrix}
&\dot{r}&\\
&\dot{p}&
\end{pmatrix}=\begin{pmatrix}
&\nu\text{e}^{\nu t}p+\text{e}^{\nu t}r&\\
&\text{e}^{\nu t}p&
\end{pmatrix}
\end{align}
and 
\begin{align}\label{s5.3}
\begin{split}
K=
\begin{pmatrix}
0 &-\text{e}^{\nu t}\\
\text{e}^{\nu t} &0
\end{pmatrix},\quad\quad\quad K^{-1}
\begin{pmatrix}
0 &\text{e}^{-\nu t}\\
-\text{e}^{-\nu t} &0 
\end{pmatrix},\\
F=\begin{pmatrix}
& \frac{1}{2}\text{e}^{\nu t}p\\
&-\frac{1}{2}\text{e}^{\nu t}r\
\end{pmatrix},\quad\quad\quad B=\frac{1}{2}\text{e}^{\nu t}(r^2+rp+p^2).
\end{split}
\end{align}
\par
Take a simple transformation $\alpha$ as
\begin{align}\label{s5.4}
\begin{split}
&\widehat{R}=\text{e}^{\nu t}\widehat{p}-\text{e}^{\nu t_0}p, \quad\quad\quad\widehat{P}=\widehat{r}-r,\\
& R=\frac{1}{2}(\widehat{r}+r),\quad\quad\quad P=-\frac{1}{2}(\text{e}^{\nu t}\widehat{p}+\text{e}^{\nu t_0}p).
\end{split}
\end{align}
The Jacobian of $\alpha$ is 
\begin{align}\label{s5.5}
\alpha_*=
\begin{pmatrix}
0 &\text{e}^{\nu t}&0&-\text{e}^{\nu t_0}\\
1 &0&-1&0\\
\frac{1}{2} &0&\frac{1}{2}&0\\
0 &-\frac{1}{2}\text{e}^{\nu t}&0&-\frac{1}{2}\text{e}^{\nu t_0}
\end{pmatrix}.
\end{align}
 The inverse transformation is 
\begin{align}\label{s5.6}
\begin{split}
&\widehat{r}=\frac{1}{2}\widehat{P}+R,\quad\quad\quad \widehat{p}=\frac{1}{2}\text{e}^{-\nu t}\widehat{R}-\text{e}^{-\nu t}P,\\
&r=-\frac{1}{2}\widehat{P}+R,\quad\quad\quad p=-\frac{1}{2}\text{e}^{\nu t_0}\widehat{R}-\text{e}^{-\nu t_0}P,
\end{split}
\end{align} 
and 
\begin{align}\label{s5.7}
\alpha_*^{-1}=
\begin{pmatrix}
0 &\frac{1}{2}&1&0\\
\frac{1}{2}\text{e}^{-\nu t}&0&0&-\text{e}^{-\nu t}\\
0 &-\frac{1}{2}&1&0\\
-\frac{1}{2}\text{e}^{-\nu t_0} &0&0&-\text{e}^{-\nu t_0}
\end{pmatrix}.
\end{align}
consequently using ($\ref{s5.4}$) and ($\ref{s5.6}$) we derive   
\begin{align}\label{s5.8}
\begin{split}
\frac{\text{d}\widehat{w}}{\text{d}t}&=
\begin{pmatrix}
&\nu\text{e}^{\nu t}\widehat{p}+\text{e}^{\nu t}\dot{\widehat{p}}&\\
& \dot{\widehat{r}}&
\end{pmatrix}=\begin{pmatrix}
&-\text{e}^{\nu t}\widehat{r}&\\
&\widehat{p}&
\end{pmatrix}=\begin{pmatrix}
&-\frac{1}{2}\text{e}^{\nu t}\widehat{P}-\text{e}^{\nu t}R &\\
&\frac{1}{2}\text{e}{-\nu t}\widehat{R}-\text{e}^{-\nu t}P &
\end{pmatrix},\\
\frac{\text{d}w}{\text{d}t}&=\begin{pmatrix}
&\frac{1}{4}\text{e}^{-\nu t}\widehat{R}-\frac{1}{2}\text{e}{-\nu t}P&\\
&\frac{1}{4}\text{e}^{\nu t}\widehat{P}+\frac{1}{2}\text{e}^{\nu t}R&
\end{pmatrix}.
\end{split}
\end{align}
Simple computation attains
\begin{align}\label{s5.9}
\begin{split}
&\phi^{(0)}_{w}=\begin{pmatrix}
&\widehat{R}&\\
&\widehat{P}&
\end{pmatrix}\left.\right|_{t=t_0}
=\begin{pmatrix}
&0&\\
&0&
\end{pmatrix},\\
&\phi^{(1)}_{w}=\frac{\text{d}\widehat{w}}{\text{d}t}\left.\right|_{t=t_0}-\phi^{(0)}_{ww}\frac{\text{d}w}{\text{d}t}\left.\right|_{t=t_0}=
\begin{pmatrix}
&-\text{e}^{\nu t_0}R&\\
&-\text{e}^{-\nu t_0}P&
\end{pmatrix}.
\end{split}
\end{align}
Set
\begin{align}\label{s5.10}
\widehat{w}=\phi^{(0)}+\phi^{(1)}\tau,
\end{align}
so we get a first order scheme for the system ($\ref{s5.2}$)
\begin{align}\label{s5.11}
\begin{pmatrix}
&r_{k+1}&\\
&p_{k+1}&
\end{pmatrix}=
\begin{pmatrix}
\frac{4-\tau^2}{4+\tau^2}&\frac{4\tau}{4+\tau^2}&\\
-\frac{4\tau}{4+\tau^2}\text{e}^{-\nu\tau}&\frac{4-\tau^2}{4+\tau^2}\text{e}^{-\nu\tau}&
\end{pmatrix}\begin{pmatrix}
&r_k&\\
&p_k&
\end{pmatrix},
\end{align}  
which is $K(z,t)$-symplectic. The transition matrix is denoted by $A$, then
\begin{align}\label{s5.12}
A^{\text{T}}
\begin{pmatrix}
0&-\text{e}^{\nu t_{k+1}}&\\
\text{e}^{\nu t_{k+1}}&0
\end{pmatrix}A=
\begin{pmatrix}
0 &-\text{e}^{\nu t_k}&\\
\text{e}^{\nu t_k}&0&
\end{pmatrix}.
\end{align}
Similarly we get $\phi^{(2)}$
\begin{align}\label{s5.13}
\phi^{(2)}=\begin{pmatrix}
&-\frac{\nu}{2}\text{e}^{\nu t_0}R&\\
&\frac{\nu}{2}\text{e}^{-\nu t_0}P&
\end{pmatrix}.
\end{align}
Direct computation also gives a second order symplectic scheme for the system ($\ref{s5.2}$)
\ba\label{s5.14}
\begin{pmatrix}
&r^{k+1}&\\
&p_{k+1}&
\end{pmatrix}=\begin{pmatrix}
\frac{16-ab}{16+ab} &\frac{8a}{16+ab}&\\
-\frac{8a}{16+ab}\text{e}^{-\nu\tau}&\frac{16-ab}{16+ab}\text{e}^{-\nu\tau}&
\end{pmatrix}\begin{pmatrix}
&r_k&\\
&p_k&
\end{pmatrix},
\ea
where 
\ba\label{s5.15}
a=2\tau-\nu\tau^2,\quad\quad\quad b=2\tau+\nu\tau^2.
\ea
The transition of the scheme ($\ref{s5.14}$) also preserves  the symplecticity.
While the direct Euler center difference scheme for the equations ($\ref{s5.2}$)
 is 
\ba\label{s5.16}
\begin{pmatrix}
&r^{k+1}&\\
&p_{k+1}&
\end{pmatrix}=\begin{pmatrix}
\frac{-\tau^2+2\nu\tau+4}{\tau^2+2\nu\tau+4}&\frac{4\tau}{\tau^2+2\nu\tau+4}&\\
\frac{-4\tau}{\tau^2+2\nu\tau+4}&\frac{-\tau^2-2\nu\tau+4}{\tau^2+2\nu\tau+4}&
\end{pmatrix}\begin{pmatrix}
&r_k&\\
&p_k&
\end{pmatrix}.
\ea
which is not a symplectic scheme.

\end{document}